\documentclass{Interspeech2024}

\usepackage{cite}
\usepackage{graphicx}  %插入图片的宏包
\usepackage{float}  %设置图片浮动位置的宏包
\usepackage{subfigure}  %插入多图时用子图显示的宏包
\usepackage{multirow}
\usepackage{lipsum}
\interspeechcameraready

\title{GenDistiller: Distilling Pre-trained Language Models based on an Autoregressive Generative Model}

\name{Yingying}{Gao}
\name{Shilei}{Zhang$^{\ast}$}
\name{Chao}{Deng}
\name{Junlan}{Feng$^{\ast}$}
\address{
  China Mobile Research, China}
\email{\{gaoyingying,zhangshilei,dengchao,fengjunlan\}@chinamobile.com}


\begin{document}

\maketitle

\newcommand\blfootnote[1]{
\begingroup
\renewcommand\thefootnote{}\footnote{#1}
\addtocounter{footnote}{-1}
\endgroup
}
\blfootnote{*Corresponding Authors}

% the abstract here must exactly match the abstract entered into the paper submission system
\begin{abstract}
    
    % 1000 characters. ASCII characters only. No citations.
    Pre-trained speech language models such as HuBERT and WavLM leverage unlabeled speech data for self-supervised learning and offer powerful representations for numerous downstream tasks. Despite the success of these models, their high requirements for memory and computing resource hinder their application on resource restricted devices. Therefore, this paper introduces GenDistiller, a novel knowledge distillation framework which generates the hidden representations of the pretrained teacher model directly by a much smaller student network. The proposed method takes the previous hidden layer as history and implements a layer-by-layer prediction of the teacher model autoregressively. Experiments on SUPERB reveal the advantage of GenDistiller over the baseline distilling method without an autoregressive framework, with 33\% fewer parameters, similar time consumption and better performance on most of the SUPERB tasks. Ultimately, the proposed GenDistiller reduces the size of WavLM by 82\%.
\end{abstract}

\section{Introduction}

Pre-trained language models (PLMs) by speech data \cite{wav2vec,wavlm,hubert,speecht5} have brought significant improvement to most of speech processing tasks. However, PLMs usually suffer from their enormous parameters and long inference time. Especially when transferring to a specific domain, it is necessary to be fine-tuned together with downstream models, which further exacerbates the consumption of computing and data resource. 

Pruning \cite{prune}, quantization \cite{quantization} and knowledge distillation (KD) \cite{kd} are the common techniques for model compression, aiming at reducing model size while retaining their performance as much as possible. In this paper, we focus on compressing PLMs based on KD, which trains a smaller student network under the supervision of the relatively large teacher network and presents significant advantages over direct training the student network using hard labels \cite{kd}. 

Most of the effective KD methods \cite{distilbert,minilm,minilmv2,ernietiny,fitnets,deit} require a layer-to-layer distillation between the intermediate layers of teacher model and student network rather than only the output layers, since different layers in PLMs contain different information and it is usually necessary to deliver multiple layers to downstream tasks to achieve better performance. This type of method need a structural mapping between teacher model and student network. Although some methods have attempted to relax \cite{tinybert} or automate \cite{dynabert} this correspondence, it is still not flexible enough. The work in \cite{distilhubert} breaks this restriction through appending several prediction heads on a student network to predict the hidden representations of teacher network directly. This simple framework provides a new paradigm for layer-wise knowledge distillation, without requiring a structural mapping between teacher network and student network. However, the prediction of different layers in \cite{distilhubert} are implemented as multi-task learning, which produces different hidden layers synchronously without considering the interrelationship between them. Moreover, the parallel prediction heads limit the number of layers to be distilled otherwise the model size will increase. We propose a new hidden layer prediction framework based on the autoregressive generative model. The reasons for using generative architecture include: 1) the autoregressive generating process takes into account the interaction between the current layer and the previous layers, in line with the process of producing hidden layer representations one-by-one in the raw network; 2) the generative model does not require future inputs other than the original features, which makes it available during the inference stage.

The major contributions include:

(1) We propose a knowledge distillation method based on a generative model, which produces the hidden layers of teacher network directly and autoregressively, considering the impacts between different layers and without the future information.

(2) We introduce an output layer and a skip-connection operation to the distilling model and verify that they are essential to make the generative distiller work.

(3) We carry out the proposed algorithm on Speech processing Universal PERformance Benchmark (SUPERB) \cite{superb} tasks and successfully reduce the WavLM size by 82\%. Compared to the counterpart without a generative structure, the proposal gets 33\% fewer parameters and better performances on most of the SUPERB tasks.

%Authors should comply with the policy on pre-prints, which can be found on the conference website. Note that this policy applies not only to pre-prints (e.g., on arXiv) but also to other material being placed in the public domain that overlaps with the content of a submitted manuscript, such as blog posts. Do not make any reference to pre-print(s) -- including extended versions -- of your submitted manuscript. 
%Note that ISCA has a general policy regarding referencing publications that have not been peer-reviewed (Section \ref{section:references}). 

\section{Related Work}

\begin{figure*}[h]
  \centering
  \includegraphics[width=0.9\linewidth]{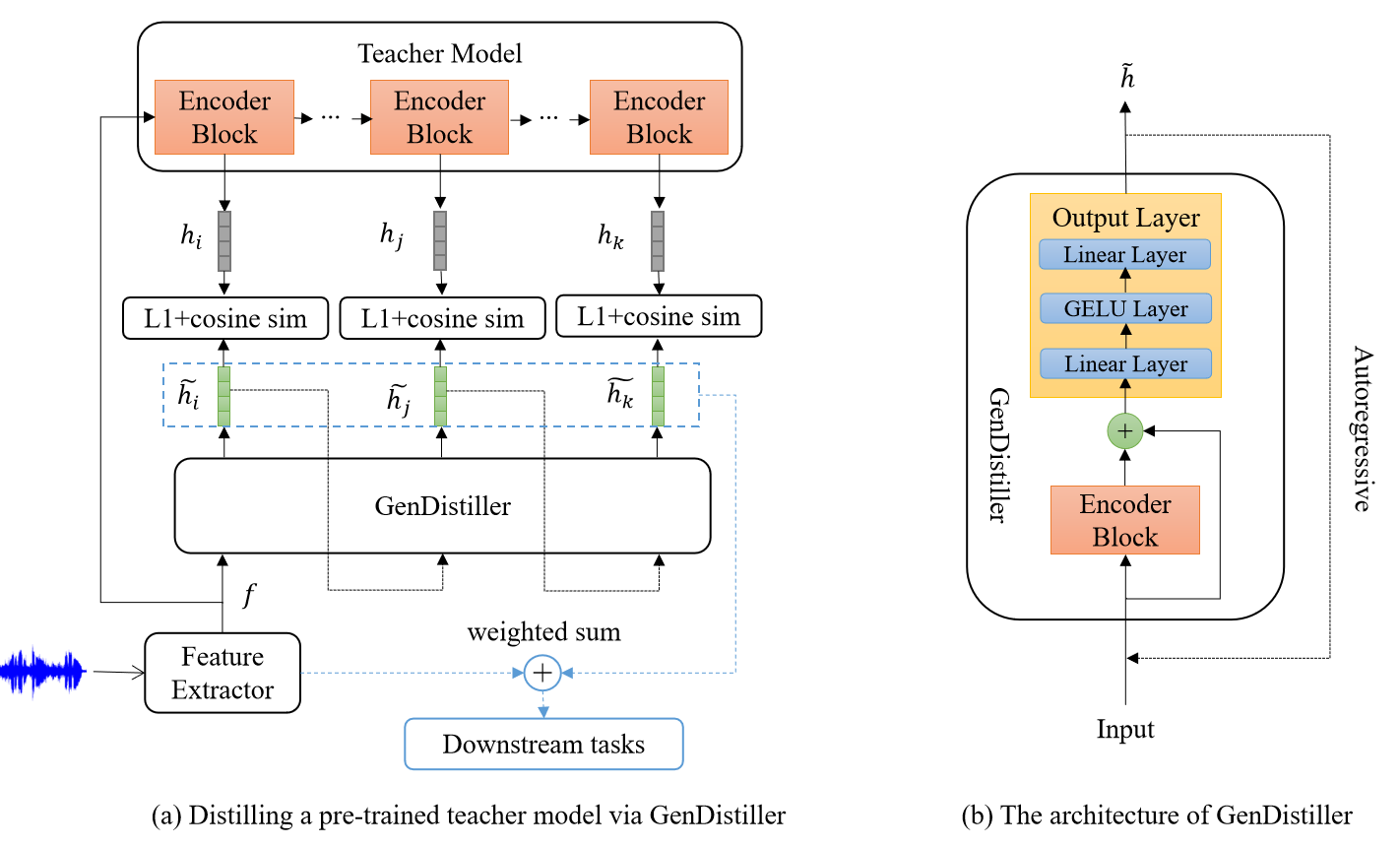}
  \caption{The distillation process and the framework of the proposed GenDistiller.}
  \label{fig1}
\end{figure*}

\subsection{Knowledge Distillation for PLMs}

Most previous work in KD \cite{tokd1,tokd2,tokd3} focuses on task-oriented distillation while it neglects the task-agnostic part which plays a crucial role for the generality of the pretrained model on downstream tasks. TinyBERT \cite{tinybert} deals with this problem through a two-stage distilling framework, including the general distillation and the task-specific distillation, which is able to transfer generic knowledge to student network and achieves a better performance on downstream tasks. However, due to the significant reduction on model size, the task-agnostic TinyBERT performs generally worse than BERT. DistilBERT \cite{distilbert} is a distilled version of BERT with a general-purpose, which inherits 97\% capabilities of the teacher model but 40\% smaller and 60\% faster. It reduces the model size of BERT through reducing the number of layers by a factor of 2 and trains the distiller by a combined loss including the distillation loss and and the language modeling loss. However, the factor-based layer reduction method restricts the structure of DistilBERT. DynaBERT \cite{dynabert} is a more flexible distiller of BERT which can adjust the width and depth to adapt different requirements for model size and latency. However, it still cannot completely avoid the layer-to-layer constraints between teacher and student networks. MINILM \cite{minilm} distills the self-attention module of the last Transformer layer which relaxes the layer mapping between teacher and student networks, and allows more flexibility for architecture of student model. MINILMv2 \cite{minilmv2} generalizes the self-attention distillation of MINILM by using multi-head self-attention relations computed by dot-product of pairs of queries, keys and values, but these type of methods are only appropriate for models based on self-attention mechanism. Some hybrid methods \cite{structprune,lighthubert,dphubert} have integrated knowledge distillation and structure pruning, and achieved very strong performance on SUPERB. Our method can also be combined with pruning approaches to seek the optimal target parameters, but this is not the focus of this work. This study mainly focuses on whether generative architecture is more helpful for generating the representaions of teacher model via a smaller student network. %In addition, the methods mentioned above are mainly structure-dependent, while we desire to explore a more general method.

\subsection{Generative Language Models}

Language models are established to predict the embedding of the target token based on the context or the previous tokens of it. In this work, we use generative language model to predict the hidden layer embeddings of teacher model in the hope to involve the interaction of the intermediate layers and avoid utilizing the future information. Three generative architectures are considered: encoder-decoder, prefix decoder and causal decoder \cite{survey}. The encoder-decoder architecture \cite{encdec,t5,bart} consists of two stacks of Transformer blocks to construct encoder and decoder separately. The encoder encodes the input sequence into a common history and the decoder generates the target sequence based on the common history in an autoregressive way. The prefix decoder architecture \cite{palm,upalm} performs bidirectional attention over the prefix tokens and unidirectional attention on generated tokens. The causal decoder architecture \cite{gpt1,gpt2,gpt3,gpt4} only attend to the past tokens of the input through a unidirectional attention mask. In our work, we select the causal decoder architecture as the backbone of our generative distiller since it is more concise and is capable to predict the hidden layer outputs of the teacher model autoregressively.

\section{Method}

We propose a knowledge distillation method based on the decoder-only generative model, which takes the previous hidden layer as history and implements a layer-by-layer prediction of the teacher model by a much smaller student network. The distilling process is shown in Figure 1(a). Take the tranformer-based pretrained speech language model for example, we set it as the teacher model and freeze all its parameters. The raw speech is delivered into a CNN-based feature extractor which is the same as the extractor in HuBERT or WavLM, then we get the feature embedding $f$. After a forward propagation of the teacher model, each encoder block outputs a hidden layer embedding $h$. Taking several hidden layer embeddings from the teacher model as a target sequence, the proposed GenDistiller generates the target embeddings autoregressively. The feature embedding $f$ is treated as the first embedding sent into GenDistiller. Then the generated hidden layers $\widetilde{h}$ is delivered back to GenDistiller as a new history to predict the next target hidden layer. $\ell_{1}$ distance and cosine similarity are calculated between the predicted hidden layers and the target hidden layers for a knowledge transfer. After the distillation, the parameters of GenDistiller are frozen and the predicted layers $\widetilde{h}$ generated by GenDistiller and the original feature $f$ are weighted summed and delivered to downstream tasks as input features.

\subsection{Model Architecture}

The architecture of the proposed GenDistiller is shown in Figure 1(b). GenDistiller is a decoder-only network which comprises only one transformer block. Notably, the input or the history of GenDistiller is added after passing the transformer block to provide a feature compensation. This operation is the same as the skip connection in ResNet and we demonstrate its necessity for speaker-related downstream tasks in this work. In addition, another essential component for GenDistiller is a projection output layer after the transformer block. The output layer consists of a nonlinear layer with GELU activation function and two linear layers before and after it. The final output of the output layer is taken as the predicted hidden layer embedding $\widetilde{h}$:
\begin{equation}
    \widetilde{h}^{l}=O(F(\widetilde{h}^{l-1})+\widetilde{h}^{l-1})
\end{equation}
in which $\widetilde{h}^{l}$ refers to the $l_{th}$ predicted hidden layer and $\widetilde{h}^{l-1}$ is its history. $F$ denotes the transformer block and $O$ represents the output layer.

%During the distillation stage, the target sequence is not delivered to GenDistiller in parallel since the sequence is two-dimensional and the attention mechanism does not work well in this scenario. Therefore, the hidden layer is generated autoregressively in both training and inference stage. Since the structure of GenDistiller is quite simple and only two target layers are produced, the training and inference time does not increased compared to the baseline distillation method.

\subsection{Distillation Loss}

To compare with other work, the distillation loss follows the loss function of DistilHuBERT \cite{distilhubert}, which consists of two parts: $\ell_{1}$ distance $\mathcal{L}_{\ell_{1}}$ and cosine similarity $\mathcal{L}_{cos}$ between the $l_{th}$ target hidden layer $h_{t}^{(l)} $ from teacher model and the generated layer $\widetilde{h}_{t}^{(l)}$ produced by student model at time $t$. The loss functions is

\begin{equation}
  \begin{split}
    & \mathcal{L}^{(l)}=\mathcal{L}_{\ell_{1}}^{(l)}+\lambda\mathcal{L}_{cos}^{(l)} \\
    & =\sum_{t=1}^{T}\left [\frac{1}{D}\Vert \widetilde{h}_{t}^{(l)}-h_{t}^{(l)} \Vert_{1}-\lambda \log{\sigma (\cos (\widetilde{h}_{t}^{(l)},h_{t}^{(l)}))}\right ] \\
  \end{split}
\end{equation}

Where $T$ refers to the time steps. $\sigma$ is sigmoid activation and  $\lambda$ controls the contribution of the cosine similarity loss.

\begin{table*}[th]
  \caption{Results on SUPERB of the proposal and baselines. The metrics include accuracy (Acc\%) phoneme error rate (PER\%), word error rate (WER\%), maximum term weighted value (MTWV), F1 score (F1\%), concept error rate (CER\%), equal error rate (EER\%), and diarization error rate (DER\%).}
  \label{tab1}
  \centering
  \begin{tabular}{l c c c c c c c c c}
    \hline
    \multirow{2}*{\textbf{Method}} & \textbf{ASR} & \textbf{SID} & \textbf{ER}  & \textbf{Qbe} & \textbf{KS} & \textbf{PR} & \textbf{ASV}  & \textbf{SD} & \textbf{SF} \\
    \cline{2-10}
    ~ & WER$\downarrow$ & Acc$\uparrow$ & Acc$\uparrow$ & MTWV$\uparrow$ & Acc$\uparrow$ & PER$\downarrow$ & EER$\downarrow$ & DER$\downarrow$ & F1$\uparrow$/CER$\downarrow$\\
    \hline
    \textbf{Baselines} & & & & & & & & &\\
    \hline
    wav2vec \cite{superb} & $15.86$  & $56.56$ & $59.79$ & $4.85$ & $95.59$ & $31.58$ & $7.99$ & $9.90$ & $76.37/43.71$   \\ 
    HuBERT Base \cite{superb}  & $6.42$  & $81.42$ & $64.92$ & $7.36$ & $96.30$ & $5.41$ & $5.11$ & $5.88$ & $88.53/25.20$     \\  
    WavLM Base \cite{structprune}  & $6.21$  & $84.51$ & $65.94$ & $8.70$ & $96.79$ & $4.84$ & $4.69$ & $4.55$ & $89.38/22.86$    \\ 
    \hline
    \textbf{Distilled Baselines} & & & & & & & & & \\
    \hline
    DistilHuBERT \cite{distilhubert} & $13.37$  & $73.54$ & $63.02$ & $5.11$ & $95.98$ & $16.27$ & $8.55$ & $6.19$ & $82.57/35.39$  \\
    DisilWavLM \cite{structprune} & $13.24$  & $71.00$ & $63.69$ & $7.07$ & $96.40$ & $14.18$ & $8.87$ & $7.2$ & $85.27/31.80$  \\
    \hline
    \textbf{Proposed} & & & & & & & & & \\
    \hline
    GenDistiller  & $12.98$  & $72.50$ & $64.33$ & $5.41$ & $96.53$ & $13.63$ & $6.68$ & $6.83$ & $84.10/11.46$  \\
    \hline
  \end{tabular} 
\end{table*}

\section{Experiments}
\subsection{Experimental Setup}
We implement our experiments with the S3PRL (Self-Supervised Speech Pre-training and Representation Learning) toolkit \cite{s3prl1,s3prl2}. The pre-trained speech language model is first distilled, then the distiller is frozen and the generated hidden representations are weighted summed as features for downstream models. The proposed model is evaluated on SUPERB \cite{superb}, which provides 10 predefined downstream tasks as a Speech processing Universal PERformance Benchmark, including phoneme recognition (PR), keyword spotting (KS), automatic speech recognition (ASR), intent classification (IC), slot filling (SF), query by example spoken term detection (QbE), speaker identification (SID), automatic speaker verification (ASV), speaker diarization (SD) and emotion recognition (ER). The intent classification task is not executed since the required Fluent Speech Commands dataset is not available at present.

\noindent\textbf{Data}. All the 960 hours of the training data in LibriSpeech \cite{librispeech} are used for knowledge distillation. For downstream SUPERB tasks, the datasets on the official guidelines are adopted.

\noindent\textbf{Model}. We choose the WavLM base model as the teacher model, which consists of a 7-layer CNN feature extractor and a 12-layer transformer encoder. Our GenDistiller has the same feature extractor but with only one transformer block (same as the encoder block in WavLM) as the backbone. Besides, one output layer (dim = 768) is inserted after the transformer block. The input of the transformer block will be added with the output of the block and the sum is sent to the output layer to gain the predicted hidden representation.

\noindent\textbf{Pretraining}. The distillation procedure is run on a 32GB V100 GPU for 200k steps with a batch size of 24 utterances, taking 54 hours. The training hyper parameters are setting according to DistilHuBERT \cite{distilhubert}. The learning rate increases linearly to 2.0e-4 in the first 7\% steps and decreases linearly to 0 in the remaining steps. $\lambda$ is set as 1 in the loss function.

\subsection{Results}
\subsubsection{SUPERB}
Results on SUPERB are shown in Table 1. First, the comparison between the baseline PLMs and the proposed approach indicates that GenDistiller inherits most of the capabilities of its teacher, and even has superior performance to the Wav2vec 2.0 base model. Second, our model outperforms the previous distilling method that predicts the hidden layers in a multi-task way instead of autoregressive generation. This observation suggests that the generative distiller learns better hidden representaions by taking its previous neighbour as history. In addition, our model has a more significant advantage on speaker related tasks such as speaker verification (ASV) and diarization (SD).

\subsubsection{Model Size and Inference Speed}
Table 2 presents the comparison on model size and inference speed. As shown in the rightmost column, GenDistiller has similar inference speed with the previous distillation method DistilWavLM without an autoregressive structure, which both offer a 40\% speedup performed on a V100 GPU (73\% in \cite{distilhubert} by 4 CPUs). However, GenDistiller further reduces the WavLM size by 82\%, which has 33\% fewer parameters than DistilWavLM but better performances on SUPERB.

\begin{table}[h]
  \caption{Model sizes and inference time performed on a V100 GPU, by extracting features from the LibriSpeech dev-clean set with a batch size of one. Results are averaged over three runs. }
  \label{tab2}
  \centering
  \begin{tabular}{l c c}
    \hline
    \multirow{2}*{\textbf{Method}} & \textbf{\#Param.} & \textbf{Inf. time} \\
    \cline{2-3}
    ~ & Millions & seconds \\
    \hline
    WavLM & $94.68(100\%)$  & 283(1.00x)   \\ 
    DistilWavLM & $26.57(25\%)$  & 200(1.42x)  \\ 
    GenDistiller & $17.58(18\%)$  & 201(1.41x)   \\ 
    \hline
  \end{tabular} 
\end{table}

\subsubsection{Model Architecture}

To make the generative distiller work, we have experimented different model architectures. In this experiment, we take three hidden layers (4,8,12) from teacher model as target, since they achieved optimal performance on SUPERB reported in \cite{distilhubert}. The top two rows in Table 3 are the results without an autoregressive framework, which takes the feature embedding and two hidden layer embeddings from two transformer blocks in student network to generate the three target layers, or three hidden layer embeddings from three transformer blocks to approximate the three target layers synchronously. It is shown that these approaches perform poorly especially for the task related with speaker (SID). We put forward only one transformer block to generate the three target layers autoregressively (as shown in the third row of Table 3), which improves the SID performance but is still not good enough. On the base of this architecture, we try to compensate the history to prevent it being forgotten during the autoregressive generation. Taking the feature embedding as a common history, we implement a cross attention on the base of self attention and find that the cross attention manner is not beneficial for neither task. At last, we add the input to the output of transformer block (skip connect) and send the sum to an output layer to constitute the final architecture of GenDistiller, which enhances the performance of both the two types of tasks. And ablation studies further prove that the output layer is particularly crucial for SID task. The success of the output layer may be attributed to the nonlinear mapping layer inside it.

\begin{table}[h]
  \caption{Model Architecture Comparison. }
  \label{tab3}
  \centering
  \begin{tabular}{l c c}
    \hline
    \multirow{2}*{\textbf{Method}} & \textbf{ASR} & \textbf{SID} \\
    \cline{2-3}
    ~ & WER$\downarrow$ & Acc$\uparrow$ \\
    \hline
    feat \& 2 layers  $\rightarrow$ 3 layers & $14.34$  &  $42.07$  \\ 
    3 layers $\rightarrow$ 3 layers & $13.21$  & $47.91$  \\ 
    1 layer $\rightarrow$ 3 layers & $13.96$  & $49.55$   \\ 
    \quad+cross attention & $17.17$  & $48.41$  \\
    \quad+skip connect (w/o output layer) & $14.78$  & $48.76$  \\
    \quad+output layer (w/o skip connect) & $12.41$  & $65.01$  \\
    \quad+skip connect+output layer (proposed)   & $12.79$  & $69.74$ \\
    \hline
  \end{tabular} 
\end{table}

Based on the above work, we conduct fewer hidden units in the only transformer block or a smaller feature extractor (change the output channel from 512 to 256) for GenDistiller to further decrease the model size. However, Table 4 illustrates that when the number of hidden units or the output channel in the convolutional layer of feature extractor is reduced, the performance of ASR or SID or both of them declines.

\begin{table}[h]
  \caption{The performance of GenDistiller with fewer hidden units or a smaller feature extractor. }
  \label{tab4}
  \centering
  \begin{tabular}{l c c c}
    \hline
    %\multirow{2}*{\textbf{\#Units}} & \textbf{ASR} & \textbf{SID} \\
    \multirow{2}*{\textbf{Method}} & \textbf{\#Param.} & \textbf{ASR} & \textbf{SID} \\
    \cline{2-4}
    ~ & Millions & WER$\downarrow$ & Acc$\uparrow$ \\
    \hline
    %3072 (proposed) & 17.58 & $12.79$  & $69.74$  \\ %3layers
    3072 hidden units(proposed) & 17.58(18\%) & $12.98$  & $72.50$  \\ %2layers
    2048 hidden units & 16.01(17\%) & $12.98$  & $71.80$  \\ %2layers
    1536 hidden units & 15.22(16\%) & $13.37$  &  $72.16$  \\ %2layers
    256 output channel & 14.24(15\%) & $13.65$  &  $67.76$  \\ %2layers
    %2048 & & $$  &  $$  \\ %3layers
    %1536 & & $13.21$  &  $68.83$  \\ %3layers
    %smaller feat extractor & $$  &  $$  \\ 3layers
    \hline
  \end{tabular} 
\end{table}
%\vspace{-0.3cm} 

\subsubsection{Target Layer}

The selection of target layers will reflect the performance and the inference speed of the distiller, therefore, we test different combinations of target layers. Different with DistilHuBERT \cite{distilhubert} that selected three target layers (4,8,12) for the optimal performance, GenDistiller is able to achieve comparable or even better performance with only two target layers (4,8). Moreover, we explore that training the GenDistiller with three target layers but generating different numbers of layers during the inference stage to figure out the effect of the generated sequence length. The bottom three rows in Table 5 demonstrate that reducing the length of the generated sequence appropriately has a negligible impact on performance, but can improve inference speed. On the other hand, increasing the length of the generated sequence is beneficial to achieve better performance.

\begin{table}[h]
  \caption{The performance of GenDistiller with different combinations of target layers and sequence lengths. }
  \label{tab5}
  \centering
  \begin{tabular}{l c c}
    \hline
    \multirow{2}*{\textbf{Target layer}} & \textbf{ASR} & \textbf{SID} \\
    \cline{2-3}
    ~ & WER$\downarrow$ & Acc$\uparrow$ \\
    \hline
    4,8(proposed) & $12.98$  & $72.50$  \\ 
    \hline
    12 & $13.21$  &  $68.83$  \\ 
    4,12 & $13.78$  & $73.63$  \\ 
    4,8,12 & $12.79$  & $69.74$  \\ 
    4,6,8,12 & $12.27$  & $70.99$  \\ 
    \hline
    \multirow{2}*{\textbf{Gen Length}} & \textbf{ASR} & \textbf{SID} \\
    \cline{2-3}
    ~ & WER$\downarrow$ & Acc$\uparrow$ \\
    \hline
    % 1 & $15.34$  &  $72.25$  \\ %2layers
    % 2 & $12.98$  & $72.50$  \\ %2layers
    % 3 & $12.94$  & $$  \\ %2layers
    2 & $13.03$  &  $69.41$  \\ %2layers
    3 & $12.79$  & $69.74$  \\ %2layers
    4 & $12.81$  & $71.26$  \\ %2layers
    \hline
  \end{tabular} 
\end{table}
%\vspace{-0.3cm} 

% \begin{table}[h]
%   \caption{Generation length. }
%   \label{tab5}
%   \centering
%   \begin{tabular}{l c c}
%     \hline
%     \multirow{2}*{\textbf{Gen Length}} & \textbf{ASR} & \textbf{SID} \\
%     \cline{2-3}
%     ~ & WER$\downarrow$ & Acc$\uparrow$ \\
%     \hline
%     2 & $13.03$  &  $69.41$  \\ 
%     3 & $12.79$  & $69.74$  \\ 
%     4 & $12.81$  & $71.26$  \\ 
%     \hline
%   \end{tabular} 
% \end{table}

% \begin{table}[h]
%   \caption{Ablation study. }
%   \label{tab3}
%   \centering
%   \begin{tabular}{l c c}
%     \hline
%     \multirow{2}*{\textbf{Method}} & \textbf{ASR} & \textbf{SID} \\
%     \cline{2-3}
%     ~ & WER$\downarrow$ & Acc$\uparrow$ \\
%     \hline
%     w/o output layer & $14.78$  &  $48.76$  \\ 
%     w/o feature accumulation & $12.41$  & $65.01$  \\ 
%     \hline
%   \end{tabular} 
% \end{table}

\section{Conclusion}
In this work, we propose a generative knowledge distillation model for large-scale pretrained language models. The proposed GenDistiller is able to generate the target hidden layers from teacher model autoregressively, considering the interactions between hidden layers and avoiding seeing the future information. The final proposal has only 18\% of the parameters of WavLM, surpasses other KD models on most of SUPERB tasks with 33\% fewer parameters. We will apply the proposed distilling method to more networks in the future.

\bibliographystyle{IEEEtran}
\bibliography{mybib}

% Generated by IEEEtran.bst, version: 1.13 (2008/09/30)
\begin{thebibliography}{10}
\providecommand{\url}[1]{#1}
\csname url@samestyle\endcsname
\providecommand{\newblock}{\relax}
\providecommand{\bibinfo}[2]{#2}
\providecommand{\BIBentrySTDinterwordspacing}{\spaceskip=0pt\relax}
\providecommand{\BIBentryALTinterwordstretchfactor}{4}
\providecommand{\BIBentryALTinterwordspacing}{\spaceskip=\fontdimen2\font plus
\BIBentryALTinterwordstretchfactor\fontdimen3\font minus
  \fontdimen4\font\relax}
\providecommand{\BIBforeignlanguage}[2]{{%
\expandafter\ifx\csname l@#1\endcsname\relax
\typeout{** WARNING: IEEEtran.bst: No hyphenation pattern has been}%
\typeout{** loaded for the language `#1'. Using the pattern for}%
\typeout{** the default language instead.}%
\else
\language=\csname l@#1\endcsname
\fi
#2}}
\providecommand{\BIBdecl}{\relax}
\BIBdecl

\bibitem{wav2vec}
S.~Schneider, A.~Baevski, R.~Collobert, and M.~Auli, ``{wav2vec: Unsupervised
  Pre-Training for Speech Recognition},'' in \emph{Interspeech 2019}, pp.
  3465--3469.

\bibitem{wavlm}
S.~Chen, C.~Wang, Z.~Chen, Y.~Wu, S.~Liu, Z.~Chen, J.~Li, N.~Kanda
  \emph{et~al.}, ``Wavlm: Large-scale self-supervised pre-training for full
  stack speech processing,'' \emph{{IEEE} J. Sel. Top. Signal Process.},
  vol.~16, no.~6, pp. 1505--1518, 2022.

\bibitem{hubert}
W.~Hsu, B.~Bolte, Y.~H. Tsai, K.~Lakhotia, R.~Salakhutdinov, and A.~Mohamed,
  ``Hubert: Self-supervised speech representation learning by masked prediction
  of hidden units,'' \emph{{IEEE} {ACM} Trans. Audio Speech Lang. Process.},
  vol.~29, pp. 3451--3460, 2021.

\bibitem{speecht5}
J.~Ao, R.~Wang, L.~Zhou, C.~Wang, S.~Ren, Y.~Wu, S.~Liu, T.~Ko \emph{et~al.},
  ``Speecht5: Unified-modal encoder-decoder pre-training for spoken language
  processing,'' in \emph{{ACL} 2022, Dublin, Ireland, May 22-27, 2022},
  S.~Muresan, P.~Nakov, and A.~Villavicencio, Eds., pp. 5723--5738.

\bibitem{prune}
S.~Han, J.~Pool, J.~Tran, and W.~J. Dally, ``Learning both weights and
  connections for efficient neural network,'' in \emph{Neural Information
  Processing Systems}, 2015.

\bibitem{quantization}
Y.~Gong, L.~Liu, M.~Yang, and L.~D. Bourdev, ``Compressing deep convolutional
  networks using vector quantization,'' \emph{CoRR}, vol. abs/1412.6115, 2014.

\bibitem{kd}
G.~E. Hinton, O.~Vinyals, and J.~Dean, ``Distilling the knowledge in a neural
  network,'' \emph{CoRR}, vol. abs/1503.02531, 2015.

\bibitem{distilbert}
V.~Sanh, L.~Debut, J.~Chaumond, and T.~Wolf, ``Distilbert, a distilled version
  of {BERT:} smaller, faster, cheaper and lighter,'' \emph{CoRR}, vol.
  abs/1910.01108, 2019.

\bibitem{minilm}
W.~Wang, F.~Wei, L.~Dong, H.~Bao, N.~Yang, and M.~Zhou, ``Minilm: Deep
  self-attention distillation for task-agnostic compression of pre-trained
  transformers,'' in \emph{NeurIPS 2020}, H.~Larochelle, M.~Ranzato,
  R.~Hadsell, M.~Balcan, and H.~Lin, Eds.

\bibitem{minilmv2}
W.~Wang, H.~Bao, S.~Huang, L.~Dong, and F.~Wei, ``Minilmv2: Multi-head
  self-attention relation distillation for compressing pretrained
  transformers,'' in \emph{Findings of {ACL} 2021}, C.~Zong, F.~Xia, W.~Li, and
  R.~Navigli, Eds., pp. 2140--2151.

\bibitem{ernietiny}
W.~Su, X.~Chen, S.~Feng, J.~Liu, W.~Liu, Y.~Sun, H.~Tian, H.~Wu, and H.~Wang,
  ``Ernie-tiny : {A} progressive distillation framework for pretrained
  transformer compression,'' \emph{CoRR}, vol. abs/2106.02241, 2021.

\bibitem{fitnets}
A.~Romero, N.~Ballas, S.~E. Kahou, A.~Chassang, C.~Gatta, and Y.~Bengio,
  ``Fitnets: Hints for thin deep nets,'' in \emph{{ICLR} 2015,}, Y.~Bengio and
  Y.~LeCun, Eds.

\bibitem{deit}
H.~Touvron, M.~Cord, M.~Douze, F.~Massa, A.~Sablayrolles, and H.~J{\'{e}}gou,
  ``Training data-efficient image transformers {\&} distillation through
  attention,'' in \emph{{ICML} 2021}, M.~Meila and T.~Zhang, Eds., vol. 139,
  pp. 10\,347--10\,357.

\bibitem{tinybert}
X.~Jiao, Y.~Yin, L.~Shang, X.~Jiang, X.~Chen, L.~Li, F.~Wang, and Q.~Liu,
  ``Tinybert: Distilling {BERT} for natural language understanding,'' in
  \emph{Findings of the Association for Computational Linguistics: {EMNLP}
  2020,}, T.~Cohn, Y.~He, and Y.~Liu, Eds., pp. 4163--4174.

\bibitem{dynabert}
L.~Hou, Z.~Huang, L.~Shang, X.~Jiang, X.~Chen, and Q.~Liu, ``Dynabert: Dynamic
  {BERT} with adaptive width and depth,'' in \emph{NeurIPS 2020},
  H.~Larochelle, M.~Ranzato, R.~Hadsell, M.~Balcan, and H.~Lin, Eds.

\bibitem{distilhubert}
H.~Chang, S.~Yang, and H.~Lee, ``Distilhubert: Speech representation learning
  by layer-wise distillation of hidden-unit bert,'' in \emph{{ICASSP} 2022,
  Virtual and Singapore, 23-27 May 2022}, pp. 7087--7091.

\bibitem{superb}
S.~wen Yang, P.-H. Chi, Y.-S. Chuang, C.-I.~J. Lai, K.~Lakhotia, Y.~Y. Lin,
  A.~T. Liu, J.~Shi, X.~Chang, G.-T. Lin, T.-H. Huang, W.-C. Tseng, K.~tik Lee,
  D.-R. Liu, Z.~Huang, S.~Dong, S.-W. Li, S.~Watanabe, A.~Mohamed, and
  H.~yi~Lee, ``{SUPERB: Speech Processing Universal PERformance Benchmark},''
  in \emph{Interspeech 2021}, pp. 1194--1198.

\bibitem{tokd1}
L.~Zhang, Y.~Shi, Z.~Shi, K.~Ma, and C.~Bao, ``Task-oriented feature
  distillation,'' in \emph{NeurIPS 2020}, H.~Larochelle, M.~Ranzato,
  R.~Hadsell, M.~Balcan, and H.~Lin, Eds.

\bibitem{tokd2}
W.~Huang, Z.~Peng, L.~Dong, F.~Wei, J.~Jiao, and Q.~Ye, ``Generic-to-specific
  distillation of masked autoencoders,'' in \emph{{CVPR} 2023}, pp.
  15\,996--16\,005.

\bibitem{tokd3}
M.~Xue, J.~Song, X.~Wang, Y.~Chen, X.~Wang, and M.~Song, ``Kdexplainer: {A}
  task-oriented attention model for explaining knowledge distillation,'' in
  \emph{{IJCAI} 2021}, Z.~Zhou, Ed., pp. 3228--3234.

\bibitem{structprune}
H.~Wang, S.~Wang, W.-Q. Zhang, S.~Hongbin, and Y.~Wan, ``{Task-Agnostic
  Structured Pruning of Speech Representation Models},'' in \emph{Interspeech
  2023}, pp. 231--235.

\bibitem{lighthubert}
R.~Wang, Q.~Bai, J.~Ao, L.~Zhou, Z.~Xiong, Z.~Wei, Y.~Zhang, T.~Ko, and H.~Li,
  ``{LightHuBERT: Lightweight and Configurable Speech Representation Learning
  with Once-for-All Hidden-Unit BERT},'' in \emph{Interspeech 2022}, pp.
  1686--1690.

\bibitem{dphubert}
Y.~Peng, Y.~Sudo, S.~Muhammad, and S.~Watanabe, ``{DPHuBERT: Joint Distillation
  and Pruning of Self-Supervised Speech Models},'' in \emph{Interspeech 2023},
  pp. 62--66.

\bibitem{survey}
W.~X. Zhao, K.~Zhou, J.~Li, T.~Tang, X.~Wang, Y.~Hou, Y.~Min, B.~Zhang
  \emph{et~al.}, ``A survey of large language models,'' \emph{CoRR}, vol.
  abs/2303.18223, 2023.

\bibitem{encdec}
A.~Vaswani, N.~Shazeer, N.~Parmar, J.~Uszkoreit, L.~Jones, A.~N. Gomez,
  L.~Kaiser, and I.~Polosukhin, ``Attention is all you need,'' in \emph{NeurIPS
  2017}, I.~Guyon, U.~von Luxburg, S.~Bengio, H.~M. Wallach, R.~Fergus,
  S.~V.~N. Vishwanathan, and R.~Garnett, Eds., pp. 5998--6008.

\bibitem{t5}
C.~Raffel, N.~Shazeer, A.~Roberts, K.~Lee, S.~Narang, M.~Matena, Y.~Zhou,
  W.~Li, and P.~J. Liu, ``Exploring the limits of transfer learning with a
  unified text-to-text transformer,'' \emph{J. Mach. Learn. Res.}, vol.~21, pp.
  140:1--140:67, 2020.

\bibitem{bart}
M.~Lewis, Y.~Liu, N.~Goyal, M.~Ghazvininejad, A.~Mohamed, O.~Levy, V.~Stoyanov,
  and L.~Zettlemoyer, ``{BART:} denoising sequence-to-sequence pre-training for
  natural language generation, translation, and comprehension,'' in \emph{{ACL}
  2020}, D.~Jurafsky, J.~Chai, N.~Schluter, and J.~R. Tetreault, Eds., pp.
  7871--7880.

\bibitem{palm}
A.~Chowdhery, S.~Narang, J.~Devlin, M.~Bosma, G.~Mishra, A.~Roberts, P.~Barham,
  H.~W. Chung, C.~Sutton \emph{et~al.}, ``Palm: Scaling language modeling with
  pathways,'' \emph{CoRR}, vol. abs/2204.02311, 2022.

\bibitem{upalm}
Y.~Tay, J.~Wei, H.~W. Chung, V.~Q. Tran, D.~R. So, S.~Shakeri, X.~Garcia, H.~S.
  Zheng, J.~Rao, A.~Chowdhery, D.~Zhou, D.~Metzler, S.~Petrov, N.~Houlsby,
  Q.~V. Le, and M.~Dehghani, ``Transcending scaling laws with 0.1{\%} extra
  compute,'' \emph{CoRR}, vol. abs/2210.11399, 2022.

\bibitem{gpt1}
\BIBentryALTinterwordspacing
A.~Radford and K.~Narasimhan, ``Improving language understanding by generative
  pre-training,'' 2018. [Online]. Available:
  \url{https://api.semanticscholar.org/CorpusID:49313245}
\BIBentrySTDinterwordspacing

\bibitem{gpt2}
\BIBentryALTinterwordspacing
A.~Radford, J.~Wu, R.~Child, D.~Luan, D.~Amodei, and I.~Sutskever, ``Language
  models are unsupervised multitask learners,'' 2019. [Online]. Available:
  \url{https://api.semanticscholar.org/CorpusID:160025533}
\BIBentrySTDinterwordspacing

\bibitem{gpt3}
T.~B. Brown, B.~Mann, N.~Ryder, M.~Subbiah, J.~Kaplan, P.~Dhariwal,
  A.~Neelakantan, P.~Shyam \emph{et~al.}, ``Language models are few-shot
  learners,'' in \emph{NeurIPS 2020}, H.~Larochelle, M.~Ranzato, R.~Hadsell,
  M.~Balcan, and H.~Lin, Eds.

\bibitem{gpt4}
\BIBentryALTinterwordspacing
OpenAI, ``{GPT-4} technical report,'' \emph{CoRR}, vol. abs/2303.08774, 2023.
  [Online]. Available: \url{https://doi.org/10.48550/arXiv.2303.08774}
\BIBentrySTDinterwordspacing

\bibitem{s3prl1}
A.~T. Liu, S.-W. Li, and H.~yi~Lee, ``Tera: Self-supervised learning of
  transformer encoder representation for speech,'' 2020.

\bibitem{s3prl2}
A.~T. Liu, S.-w. Yang, P.-H. Chi, P.-c. Hsu, and H.-y. Lee, ``Mockingjay:
  Unsupervised speech representation learning with deep bidirectional
  transformer encoders,'' \emph{ICASSP 2020}.

\bibitem{librispeech}
V.~Panayotov, G.~Chen, D.~Povey, and S.~Khudanpur, ``Librispeech: an asr corpus
  based on public domain audio books,'' in \emph{ICASSP 2015}, pp. 5206--5210.

\end{thebibliography}

\end{document}